\documentclass[a4paper]{jpconf}
\usepackage{amsmath,graphicx}

\begin{document}
\title{$v_4$ for identified particles at RHIC from viscous hydrodynamics}
\author{Matthew Luzum}
\address{Institut de physique th\'eorique,
CEA Saclay, 91191 Gif-sur-Yvette, France}
\ead{matthew.luzum@cea.fr}
\begin{abstract}
Using ideal and viscous hydrodynamics, the ratio of azimuthal moments $v_4/(v_2)^2$ is calculated for pions, protons, and kaons in $\sqrt{s}$=200 A*GeV Au+Au collisions.   For any value of viscosity here is little dependence on particle species.  Ideal hydrodynamics and data show a flat curve as a function of $p_t$.  
Adding viscosity in the standard way destroys this flatness.  However, it can be restored by replacing the standard quadratic ansatz for $\delta f$ (the viscous correction to the distribution function at freeze-out) with a weaker momentum dependence.
\end{abstract}
\section{Introduction}
The azimuthal distribution of detected particles in a relativistic heavy ion collision contains valuable information about the medium created in such a collision.  This angular distribution is typically characterized by its Fourier components
\begin{equation}
\label{fourier}
E\frac{d^3 N}{d^3 {\bf p}}
= v_0 \left[1 + \sum_{n = 1}^\infty 2 v_n\, \cos(n\, \phi) \right],
\end{equation}
where $\phi$ is the angle of the outgoing particle momentum with respect to the collision plane for each event.

The first non-vanishing coefficient is the elliptic flow coefficient $v_2$, which is an important observable that is sensitive to bulk properties of the collision fireball and has been the object of much study~\cite{Voloshin:2008dg}.  Much less studied is the next coefficient, $v_4$.  In these proceedings, results are presented from ideal and viscous hydrodynamic simulations of Au+Au collisions at the Relativistic Heavy Ion Collider (RHIC).  The main results of these calculations can be found in Ref.~\cite{Luzum:2010ae}, to which this can be considered a companion article.  The details of the viscous hydrodynamic model can be found in Ref.~\cite{Luzum:2008cw}.
\section{Analytic prediction}
A simple argument by Borghini and Ollitrault \cite{Borghini:2005kd} indicates that $v_4$ is expected, at large $p_t$, to be largely determined by $v_2$.  It is then most useful to construct the ratio $v_4/(v_2)^2$ and investigate deviations from this ``ideal hydrodynamic'' scaling.

To summarize, one can define a function $u_{\rm max}(\phi)$, which is the magnitude of maximum fluid velocity on the   freeze-out surface which is flowing in the azimuthal direction $\phi$.  For symmetric collision systems, this fluid velocity on the   freeze-out surface will have only even Fourier components:
\begin{equation}
u_{\rm max}(\phi) = U(1 + 2V_2 \cos(2\phi) + 2V_4 \cos(4\phi) + \ldots )\, .
\end{equation}
If one plugs this expression into the ideal (zero viscosity) Cooper-Frye formula (neglecting quantum statistics)
\begin{equation}
\label{CF}
E\frac{d^3 N}{d^3 {\bf p}}  
\propto \int p_{\mu} d\Sigma^\mu \exp \left(- \frac {p_\mu u^\mu} {T} \right)\ ,
\end{equation}
performs a saddle-point integration, and expands to leading order in these ``intrinsic'' elliptic and quadrangular flow coefficients $V_2\ll1$ and $V_4\ll1$,  the result is
\begin{align}
v_2(p_t)=&\frac{V_2 U}{T}\left(p_t-m_t v\right)\\
v_4(p_t)=&\frac{1}{2}\frac{(V_2 U)^2}{T^2}\left(p_t-m_t v\right)^2+\frac{V_4 U}{T}\left(p_t-m_t v\right) \nonumber \\
\label{v4}
=&\frac{1}{2}v_2(p_t)^2 + \frac {V_4} {V_2} v_2(p_t),
\end{align}
with $m_t=\sqrt{p_t^2+m^2}$ and  $v\equiv U/\sqrt{1+U^2}$.  This suggests that it is useful to construct the ratio $v_4/(v_2)^2$:
\begin{equation}
\label{ratio}
\frac {v_4} {(v_2)^2} =\frac 1 2 + \frac {V_4 T} {V_2 U\left(p_t-m_t v\right)}  = \frac 1 2 + \frac {V_4} {V_2} \frac {1} {v_2(p_t)}.
\end{equation}
\begin{figure}[b]
\begin{center}
\includegraphics[width=0.7\linewidth]{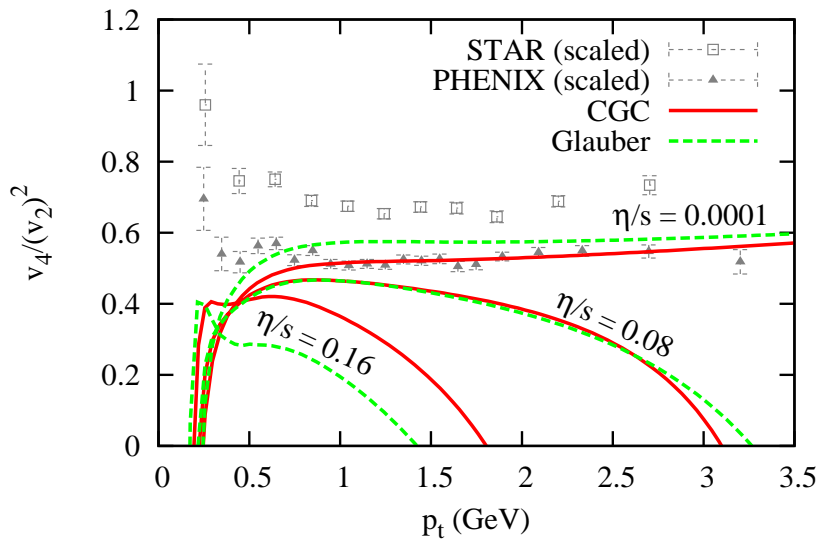}
\end{center}
\caption{\label{data}Charged hadron $v_4/(v_2)^2$ as a function of transverse momentum, using Glauber and CGC initial conditions with impact parameter $b = 10$ fm, and using the standard quadratic ansatz for $\delta f$.  Data are from STAR (40-50\% centrality)~\cite{YutingPhD} and PHENIX (45-50\% centrality)~\cite{Adare:2010ux}, scaled down by a factor of 1.38 to account for fluctuations~\cite{Gombeaud:2009ye,Luzum:2010ae}.}
\end{figure}
So, according to this prediction, the ratio should approach 1/2 at large $p_t$, with 1/$p_t$ corrections at smaller transverse momenta.  Note that this prediction does not depend on any details of the collision, as long as the assumptions are satisfied (that $V_4$ and $V_2$ are small and that one is considering large enough $p_t$ compared to the freeze out temperature).   Also note that, when plotted as a function of $v_2$ instead of $p_t$, there should be no dependence on the particle species.
\section{Results}
Results from ideal hydrodynamics confirm these expectations.  As detailed in Ref.~\cite{Luzum:2010ae}, when the viscosity is set to zero, the ratio $v_4/(v_2)^2$ approaches roughly 1/2 at large $p_t$ with a correction that behaves as 1/$p_t$, as predicted from Eq.~\eqref{ratio}.  Both the magnitude and sign of this correction depend strongly on the freeze-out temperature.   A freeze-out temperature close to $T_f = 140 MeV$ (the temperature that gives the best fit to other data~\cite{Luzum:2008cw}) gives a flat dependence on \(p_t\), which is what is seen experimentally (see, e.g., Fig.~\ref{data}).  

Using different choices of initial conditions for hydrodynamic evolution [Glauber-type or Color-Glass-Condensate (CGC)-type] has a small effect, indicating little sensitivity to the initial eccentricity.  This is in contrast to $v_2$ itself, which is largely driven by the initial eccentricity.  

As predicted, and as seen experimentally~\cite{Huang:2008vd,Huang:2009zzh}, there is no dependence on particle species at high $p_t$.  Since there is little dependence on transverse momentum, this is true whether $v_4/(v_2)^2$ is plotted versus $p_t$ or $v_2$. 

Data, however, show a value that is closer to 1, for a large range of centrality and $p_t$.  Most of the discrepancy can be understood by noting that $v_4$ and $v_2$ are obtained separately when averaged over events.  Event-by-event fluctuations in $v_4$ and $v_2$, which are not included in these hydrodynamic calculations, then increase the ratio $v_4/(v_2)^2$~\cite{Gombeaud:2009ye}.
%
%
%
%
\begin{figure}[b]
\begin{center}
\includegraphics[width=0.7\linewidth]{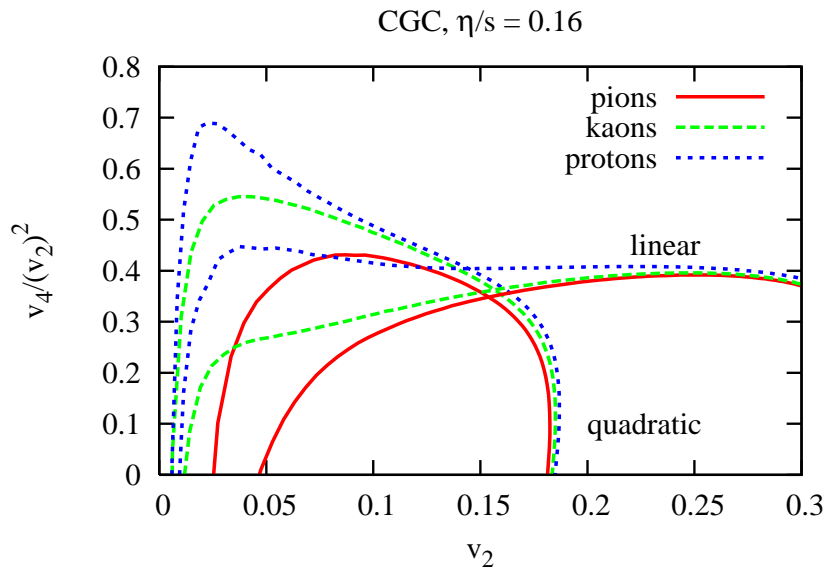}
\end{center}
\caption{\label{identified}Identified particle $v_4/(v_2)^2$ versus $v_2$ from viscous hydrodynamics with CGC initial conditions and $\eta/s = 0.16$ at $b = 8$ fm using the standard quadratic ansatz as well as a linear ansatz for the viscous correction, $\delta f$, to the equilibrium distribution function at freeze-out~\cite{Luzum:2010ad}. 
}
\end{figure}

To add viscosity to the calculation, one must first make a choice for \(\delta f\), the viscous correction to the equilibrium distribution function at freeze-out.  Until recently, all groups used a quadratic ansatz for the momentum dependence of \(\delta f\), but the correct form is not known, and it can have a significant effect~\cite{Luzum:2010ad}.

Adding viscosity and using the standard quadratic ansatz results in a significant change to the $p_t$ dependence of $v_4/(v_2)^2$, as shown in Fig.~\ref{data}.  There is again little dependence on the initial conditions.  Although there is some dependence of the overall size of $v_4/(v_2)^2$ on initial conditions, the shape of the curve is largely determined by only the viscosity.  In addition, a large centrality dependence is introduced (not shown).  Figure~\ref{identified} shows that, as with ideal hydrodynamics results,  there is little dependence on particle species at high $p_t$ when plotted as a function of $v_2$, even when the viscous corrections are large, and regardless of the choice for $\delta f$.

Using a weaker momentum dependence for $\delta f$ can make the $p_t$ dependence of $v_4/(v_2)^2$ more flat (see Fig.~\ref{identified}), and at the same time reduce the dependence on impact parameter.  This may indicate that the standard quadratic ansatz is not correct and could give insight into the behavior of the hadron gas present at freeze-out (see Ref.~\cite{Luzum:2010ad} for more details).
\section{Conclusions}
Viscous hydrodynamic simulations of heavy ion collisions were performed, and the ratio of azimuthal moments $v_4/(v_2)^2$ was computed.  It was found that, as in ideal hydrodynamics, identified particles in viscous hydrodynamic simulations have the same value, except at low $p_t$, when plotted as a function of $v_2$.  In addition, the shape as a function of $p_t$ depends little on the initial conditions, but depends strongly on the shear viscosity to entropy density ratio $\eta/s$, when implemented using the standard quadratic ansatz for the viscous correction to the equilibrium distribution function at freeze-out.   In contrast, a weaker momentum dependence for $\delta f$ can restore the flat dependence on transverse momentum that is seen in ideal hydrodynamics and data, and reduce the strong centrality dependence, which is not seen in experiment.
%
%
%
%
%
%
%
%
%
%
%
%
%
%
\ack
The author would like to thank collaborators Jean-Yves Ollitrault and Cl\'ement Gombeaud, as well as Paul Romatschke for the original version of the viscous hydro code.  This work was funded by ``Agence Nationale de la Recherche'' under grant
ANR-08-BLAN-0093-01.
\section*{References}
\bibliographystyle{iopart-num}
\bibliography{HotQuarks}
\end{document}